\documentclass[12pt,a4paper]{article}
\usepackage{mathtools}
\usepackage[sc,noBBpl]{mathpazo}
\usepackage[mathscr]{eucal}
\usepackage{amssymb,amsmath}
\usepackage{amsfonts,amsthm}
\usepackage{graphicx}
\usepackage{subfigure}
\usepackage{multirow}
\usepackage[sc,noBBpl]{mathpazo}
\usepackage[mathscr]{eucal}
\usepackage{booktabs}
\usepackage[utf8]{inputenc}
\usepackage{tgpagella}
\usepackage{array}
\theoremstyle{plain}
\usepackage{hyperref}
\newcommand\scI{{\mathscr I}}

\newcommand\scN{{\mathscr N}}

\newcommand\mvector{\boldsymbol}

\newcommand\vc{\mvector{c}}
\newcommand\vd{\mvector{d}}

\newcommand\vp{\mvector{p}}
\newcommand\vq{\mvector{q}}

\newcommand\vC{\mvector{C}}

\newcommand\vGamma{\mvector{\Gamma}}

\newcommand\field{\mathbb}

\newcommand\R{\field{R}}

\newcommand\CC{\field{C}}
\newcommand\Z{\field{Z}}

\newcommand\Dt{\frac{\mathrm{d}\phantom{t} }{\mathrm{d}\mspace{1mu}
t}}

\newcommand\oder[2]{\dfrac{\mathrm{d} #1 }{\mathrm{d} #2}}
\newcommand\pder[2]{\dfrac{\partial #1 }{\partial #2}}

\newcommand\defset[2]{\left\{{#1}\;\Big\vert \;\; {#2} \,\right\}}

\newtheorem{theorem}{Theorem}

\newtheoremstyle{note}{\topsep}{\topsep}{\slshape}{}{\scshape}{}{ }{}
\theoremstyle{note}

\numberwithin{equation}{section}
\numberwithin{theorem}{section}
\numberwithin{definition}{section}
\numberwithin{lemma}{section}
\numberwithin{proposition}{section}
\numberwithin{corollary}{section}
\numberwithin{remark}{section}
\usepackage{xcolor}
\hypersetup{
    colorlinks,
    linkcolor={red!60!black},
    citecolor={blue!50!black},
    urlcolor={blue!80!black}
}
\title{Note on integrability of certain homogeneous Hamiltonian systems}
\author{
   Wojciech Szumi\'nski$^1$, Andrzej J.~Maciejewski$^2$  \\ 
   and  Maria Przybylska$^1$ \\[1em]
  {}$^{1}$Institute of Physics, \\ University of Zielona G\'ora, 
  Licealna 9,  \\
  PL-65-407,  Zielona G\'ora, Poland\\[1em]
   {}$^2$Institute of Astronomy,\\ University of Zielona G\'ora, 
  Licealna 9,  \\
  PL-65-407,  Zielona G\'ora, Poland }
\begin{document}
%
\maketitle
\begin{abstract}
In this paper we investigate a class of natural Hamiltonian systems
with two degrees of freedom.  The kinetic energy depends on coordinates
but the system is homogeneous.  Thanks to this property it admits, in
a general case, a particular solution. Using this solution we derive necessary
conditions for the integrability of such systems investigating
differential Galois group of variational equations.   
\end{abstract}
\date{\small Key words: integrability obstructions; Liouville integrability; differential Galois theory;  systems in polar coordinates; systems in curved spaces }
\maketitle
\mathtoolsset{%
mathic,centercolon%
}
%

\section{Introduction}

It seems that the most effective methods of proving non-integrability are based
on application of the differential Galois theory.  For Hamiltonian systems
necessary conditions for the integrability in the Liouville sense are given by the Morales-Ramis
theorem.
\begin{theorem}[Morales-Ruiz and Ramis]
  \label{th:Morales}
  Assume that a Hamiltonian system is meromorphically integrable in the
  Liouville sense in a neighbourhood of a phase curve $\vGamma$ corresponding to
  a particular solution. Then, the identity component $\mathcal{G}^0$ of the
  differential Galois group $\mathcal{G}$ of variational equations along
  $\vGamma$ is Abelian.
\end{theorem}
For a detailed exposition and a proof see e.g.~\cite{Morales:99::,Morales:01::a}.

The above theorem has found a very effective application for natural systems
given by the following Hamiltonian
\begin{equation}
  \label{eq:1}
  H = \frac{1}{2}\sum_{i=1}^n p_i^2 +V (\vq),
\end{equation}
where $V(\vq)$ is a homogeneous function of degree $k\in\Z$, and
$\vq=(q_1, \ldots, q_n)$ and $\vp=(p_1, \ldots, p_n)$ are the generalised
coordinates and momenta, respectively.  Let us note that for application of
Theorem~\ref{th:Morales} we have to know a particular solution of the considered
system. In general it is a difficult problem how to find such a
solution. However for systems given by \eqref{eq:1} with a homogeneous potential
$V (\vq)$ it is well known that if $\vd\in\CC^n$ is a non-zero solution of
nonlinear system $V'(\vd)=\vd$, then functions
\begin{equation}
  \label{eq:2}
  \vq(t) = \varphi(t) \vd, \qquad \vp(t)= \varphi(t) \vd, \qquad \ddot
  \varphi = -\varphi ^{k-1},
\end{equation} 
determine a particular solution of Hamilton's equations.  The variational
equations along this solution split into a direct product of second order
equations of the form
\begin{equation}
  \label{eq:3}
  \ddot x = -\lambda \varphi(t) ^{k-2} x,
\end{equation}
where $\lambda$ is an eigenvalue of Hessian $V''(\vd)$.  The necessary
conditions for the integrability have the form of arithmetic restrictions on
$\lambda$, see e.g.~\cite{Morales:99::,Morales:01::a}.  The crucial role in
derivation of these conditions plays the Yoshida change of independent variable
which transforms equation~\eqref{eq:3} into the Gauss hypergeometric
equation~\cite{Yoshida:87::}.

Hamiltonian~\eqref{eq:1} describes a particle moving under influence of
potential forces in flat Euclidean space $\R^n$.  It is a natural to ask what is
an analog of homogeneous systems in curved spaces.  There is no obvious answer
to this question. We have to take into account the form of metric of the
configuration space as well as the form of the potential.  We leave a general
discussion of this problem to a separate paper and here we consider systems with
two degrees of freedom given by the following Hamiltonian
\begin{equation}
  \label{eq:m_h}
  H=T+V,\qquad
  T=\frac{1}{2}r^{m-k}\left(p_r^2+\frac{p_\varphi^2}{r^2}\right),
  \qquad	 V=r^m U(\varphi),
\end{equation}
where $m$ and $k$ are integers, and $k\neq 0$.  If we consider $(r,\varphi)$ as
the polar coordinates, then the kinetic energy corresponds to a singular metric
on a plane or a sphere.  We assume that $U(\varphi)$ is a complex meromorphic
function of variable $\varphi\in\CC$, and we do not require that $U(\varphi)$ is
periodic.

The main result of this paper is the following theorem which gives necessary
conditions for the integrability of Hamiltonian systems given
by~\eqref{eq:m_h}. For its formulation we need to define the following sets
\begin{align}
  \label{eq:40}
  \scI_0(k,m):= &\defset{\dfrac{1}{k} \left( mp+1 \right) \left( 2mp+k
                  \right) }{p\in \Z}, \\
  \scI_1(k,m):= &\defset{\dfrac{1}{2k} \left( m p-2 \right) \left( mp-k
                  \right) }{p=2r+1, r\in \Z}, \\
  \scI_2(k,m):=&\defset{ \dfrac{1}{8k}  \left[ 4  m^2 \left( p
                 +\dfrac{1}{2}\right)^2 -(k-2)^{2}\right]}{p\in \Z},\\
  \scI_3(k,m):=&\defset{ \dfrac{1}{8k}  \left[  4 m^2 \left(  p
                 +\dfrac{1}{3}\right)^2 -(k-2)^{2}\right]}{p\in \Z},\\
  \scI_4(k,m):=&\defset{ \dfrac{1}{8k}  \left[  4 m^2 \left(  p
                 +\dfrac{1}{4}\right)^2 -(k-2)^{2}\right]}{p\in \Z},\\
  \scI_5(k,m):=&\defset{ \dfrac{1}{8k}  \left[  4 m^2 \left(  p
                 +\dfrac{1}{5}\right)^2 -(k-2)^{2}\right]}{p\in \Z},\\
  \scI_6(k,m):=&\defset{ \dfrac{1}{8k}  \left[  4 m^2 \left(  p
                 +\dfrac{2}{5}\right)^2 -(k-2)^{2}\right]}{p\in \Z},
\end{align}
and we put
\begin{equation}
  \label{eq:6}
  \scI_{\mathrm{a}}(k,m):=  \scI_0(k,m)\cup  \scI_1(k,m) \cup  \scI_2(k,m).
\end{equation}
\begin{theorem}
  \label{th:1}
  Assume that $U(\varphi)$ is a complex meromorphic function and  there
  exists $\varphi_0\in\CC$ such that $U'(\varphi_0)=0$ and $U(\varphi_0)\neq 0$.
  If the Hamiltonian system defined by Hamiltonian~\eqref{eq:m_h} is integrable
  in the Liouville sense, then number
  \begin{equation}
    \label{eq:4}
    \lambda := 1 + \frac{U''(\varphi_0)}{k U(\varphi_0)}, 
  \end{equation} 
  belongs to set $\scI(k,m)$ which is defined by the following table
  \begin{table}[h]
    \begin{center}
      \begin{tabular}{llll}
        \toprule
        \text{No.} & $k$ & $m$ & $\scI(k,m)$ \\  
        \midrule
        1 & $k=-2(mp+1)$ & $m$ & $\CC$ \\ 
        2 &  $k\in\Z\setminus\{0\}$ &  $m$ & $\scI_{\mathrm{a}}(k,m)$ \\ 
        \addlinespace[0.5em]
        3 & $k = 2(m p-1) \pm \frac{1}{3}m$ & $3 q$&  $\bigcup_{i=0}^6\scI_i(k,m)$\\
        \addlinespace[0.5em]
        4 & $k = 2(m p-1) \pm \frac{1}{2}m$ & $2 q$ & $\scI_{\mathrm{a}}(k,m) \cup
                                                      \scI_4(k,m)$ \\ 
        \addlinespace[0.5em]
        5 &  $k = 2(m p-1) \pm \frac{3}{5}m$  & $5q$ & $\scI_{\mathrm{a}}(k,m) \cup
                                                       \scI_3(k,m) \cup
                                                       \scI_6(k,m)$ \\
        \addlinespace[0.5em]
        6 & $k = 2(m p-1) \pm \frac{1}{5}m$  & $5q$ &  $\scI_{\mathrm{a}}(k,m) \cup
                                                      \scI_3(k,m) \cup
                                                      \scI_5(k,m)$ \\
        \addlinespace[0.5em]
        \bottomrule
      \end{tabular}
    \end{center}
    \caption{Integrability table. Here $k,m,p,q\in\Z$ and $k\neq 0$. \label{tab:integrability_table}}
  \end{table}
\end{theorem}
The above theorem tells us that if  $k=-2(mp+1)$,  then the Morales-Ramis
Theorem~\ref{th:Morales} does not give any obstruction for the integrability of
the considered systems.  Let us notice that this is an infinite family of
systems.   For systems~\eqref{eq:1} with homogeneous potentials  only two 
cases of this type are such distinguished, namely   $k=\pm 2$~\cite{Morales:99::,Morales:01::a}.   

For each pair $(k,m)$ of integers    which do not satisfy relation
$k=-2(pm+1)$, $p\in\Z$,  Theorem~\ref{th:1} restricts admissible values  $\lambda$  to the
set $ \scI_{\mathrm{a}}(k,m)$. If $m$ is not a multiple of $2$, $3$,  and
$5$ these are the only restrictions.   Otherwise, if $m$ is a multiple of
$q\in\{2,3,5\}$, and $k$ takes appropriate value, then the set of admissible
values of $\lambda$ contains additional elements. These are Cases 3--6 in
Table~\ref{tab:integrability_table}.

Let us note that the above theorem remains valid for rational $k$ and $m$.  In
such extended version we require that $k$ is a non-zero rational number, and the
restriction contained in the third column of Table~\ref{tab:integrability_table}
can be ignored.  For the proof of this extended version one has to apply a
reasoning similar to that one used in~\cite{MR3047475}.

Let us remark that there is also other possibility  to generalise systems  given
by~\eqref{eq:1} with
homogeneous potentials in such a way that  they will admit a straight line
particular solution and the variational equations can be reduced to a direct product of  hypergeometric equations. In \cite{Nakagawa:01::} the authors consider system
with Hamiltonian 
\begin{equation}
\label{eq:18}
H=T(\vp) + V(\vq).
\end{equation}
where $T $ and $V$ are homogeneous functions of integer degrees.  To find a straight line particular
solution one must solve overdetermined system of nonlinear equations 
\begin{equation*}
T'(\vc) =\vc, \qquad V'(\vc)=\vc,
\end{equation*}
that has a solution only in special cases. Moreover, this generalisation does not have a form of a natural Hamiltonian system. In other words, except the case when $\deg T =2$, it cannot be considered as a Hamiltonian function of a point in a curved space.

\section{Proof of Theorem~\ref{th:1}}

Equations of motion corresponding to Hamiltonian~\eqref{eq:m_h} have
the form
\begin{equation}
  \label{eq:m_vh}
  \begin{split} 
    & \dot r =\pder{H}{p_r}=r^{m-k}p_r,\\
    & \dot
    p_r=-\pder{H}{r}=r^{m-k-3}p_\varphi^2-\frac{1}{2}(m-k)r^{m-k-1}\left(p_r^2+\frac{p_\varphi^2}{r^2}\right)
    -mr^{m-1}U(\varphi),\\
    &\dot \varphi=\pder{H}{p_{\varphi}}=r^{m-k-2}p_\varphi ,\\
    &\dot p_\varphi=-\pder{H}{\varphi}=-r^m U'(\varphi).
  \end{split}
\end{equation}
If $U'(\varphi_0)=0$ for a certain $\varphi_0\in\CC$, then
system~\eqref{eq:m_vh} has two dimensional invariant manifold
\begin{equation}
  \label{eq:vh0_N}
  \scN=\left\{(r, p_r, \varphi,p_\varphi )\in \CC^4|\varphi=\varphi_0, \     p_\varphi=0\right\}.
\end{equation}
Indeed, equations~\eqref{eq:m_vh} restricted to $\scN$ read
\begin{equation}
  \label{eq:m_vh0}
  \dot r=r^{m-k}p_r,\qquad \dot p_r=-\frac{1}{2}(m-k)r^{m-k-1}p_r^2-mr^{m-1}U(\varphi_0).
\end{equation}
Hence, $\scN$ is foliated by phase curves parametrised by energy $E$
\begin{equation}
  \label{eq:5}
  E = \frac{1}{2}r^{m-k} p_r^2 +r^m U(\varphi_0).
\end{equation}
Taking into account that $\dot r = r^{m-k} p_r$ we can rewrite
equation~\eqref{eq:5} in the form
\begin{equation}
  \label{eq:m_H20}
  \dot r^2=2r^{m-k}\left\{E-r^mU(\varphi_0)\right\}.
\end{equation}

Let $[R,P_R, \Phi,P_{\Phi}]^T$ denote the variations of
$[r,p_r,\varphi,p_{\varphi}]^T$. Then, the variational equations along a
particular solution lying on $\scN$ take the form
\begin{equation}
  \label{eq:m_var}
\Dt
  \begin{bmatrix}
    R \\  P_R \\ \Phi  \\  P_\Phi
  \end{bmatrix}= \vC \begin{bmatrix}
    R \\  P_R \\ \Phi  \\  P_\Phi
  \end{bmatrix},
\end{equation}
with
\[
\vC=
\begin{bmatrix}
  l  r^{l-1}p_r & r^l & 0 & 0\\
 -\frac{1}{2} (l-1) l r^{l-2}p_r^2 - (m-1) m
  r^{m-2}U(\varphi_0) & -l
  r^{l-1}p_r & 0 & 0\\
  0 & 0 & 0 & r^{l-2} \\
  0 &  0 &- r^m U''(\varphi_0) & 0 
\end{bmatrix},
\]
where we introduced auxiliary parameter $l=m-k$.  Equations for $\Phi$ and $P_\Phi$ form a closed
subsystem which is called normal variational equations. This system
can be rewritten as a one second-order differential equation
\begin{equation}
  \ddot \Phi+P\dot \Phi + Q\Phi=0,\quad P=(k-m+2)r^{m-k-1}p_r,\quad Q=r^{2m-k-2}U''(\phi_0).
  \label{eq:m_eqimeqt}
\end{equation}
In order to rationalise it we make the transformation
\begin{equation}
  \label{eq:m_zz}
  t\longrightarrow z=\frac{U(\varphi_0)}{E}r^m(t),
\end{equation}
for $E\neq 0$, that gives immediately
\[
\dot z^2=-2Em^2r^{m-k-2}z^2(z-1),\quad \ddot
z=Emr^{m-k-2}z\left[(k-4m+2)z+3m-k-2\right].
\]
Equation~\eqref{eq:m_eqimeqt} after such a change of independent
variable takes the form
\begin{equation}
  \label{eq:m_eqq}
 z(z-1)\Phi''(z)+\left[\frac{2m+k+2}{2m}z-\frac{k+m+2}{2m}\right]\Phi'(z)+\frac{k(1-\lambda)}{2m^2}\Phi(z)=0, 
\end{equation}
where prime denotes derivative with respect to $z$ and
\[
\lambda = 1+\dfrac{U''(\varphi_0)}{kU(\varphi_0)}.
\]
Equation~\eqref{eq:m_eqq}
is a special case of the  Gauss hypergeometric differential
equation whose  general form is the following 
\begin{equation}
  \label{eq:gauss_hyper}
  z(z-1) \Phi''(z)+\left[(\alpha+\beta+1)z-\gamma\right]\Phi'(z)+\alpha\beta\Phi(z)=0,
\end{equation}
and $\alpha, \beta$ and $\gamma$ are parameters, see e.g.~\cite{Whittaker:35::,Kristensson:12::}.
In our case the parameters 
take the forms
\begin{equation}
  \label{eq:parameters}
  \alpha=\frac{k+2-\Delta }{4m},\qquad 
  \beta=\frac{k+2+\Delta }{4m},\qquad \gamma=\frac{k+2 +m}{2m},
\end{equation}
where
\begin{equation*}
  \Delta= \sqrt{(k-2)^2+8k\lambda}.
\end{equation*}
The differences of exponents at singularities $z=0$, $z=1$ and at
$z=\infty$ are given by
\[
\rho=1-\gamma,\qquad \sigma=\gamma-\alpha-\beta,\qquad
\tau=\beta-\alpha,
\]
respectively, so in our case they are
\begin{equation}
  \label{eq:diff_of_exponents}
  \rho=\frac{m-k-2}{2m},\qquad \sigma=\frac{1}{2}, \qquad
  \tau=\frac{\Delta}{2m}.
\end{equation}%
If Hamilton equations~\eqref{eq:m_vh} are integrable in the Liouville
sense, then by Theorem~\ref{th:Morales} the identity component of the
differential Galois group of variational equations \eqref{eq:m_var} as
well as normal variational equations~\eqref{eq:m_eqq} is Abelian, so
in particular it is solvable.  Necessary and sufficient conditions for
solvability of the identity component of the differential Galois group
for the Riemann $P$ equation as well as its special form: the
hypergeometric equation are well known thanks to the Kimura theorem
which we recall in Appendix~\ref{sec:hyper}.

The proof of Theorem~\ref{th:1} consists in a direct application of
Theorem~\ref{th:Kimura} to our Gauss hypergeometric
equation~\eqref{eq:m_eqq}.

The condition A of Theorem~\ref{th:Kimura} is fulfilled if at least
one of the following numbers
\begin{align*}
  \rho+\sigma+\tau &=\frac{2m-k-2+\Delta}{2m},  \\
  -\rho+\sigma+\tau&=\frac{k+2+\Delta}{2m}, \\
  \rho -\sigma +\tau &=\frac{-k-2+\Delta}{2m}, \\
  \rho+\sigma-\tau &=\frac{2m-k-2-\Delta}{2m}
\end{align*}
is an odd integer.  If it is the first one, then
$\lambda\in\scI_0(k,m)$, and if it is the second one, then
$\lambda\in\scI_1(k,m)$.  It is easy to check that if the third or
fourth of the above numbers is an odd integer, then
$\lambda \in \scI_0(k,m) \cup \scI_1(k,m)$. This exhaust all the
possibilities in Case A of Theorem~\ref{th:Kimura}.

Now, we pass to Case B of Theorem~\ref{th:Kimura}. In this case the
quantities $\rho$ or $-\rho$, $\sigma$ or $-\sigma$ and $\tau$ or
$-\tau$ must belong to Table~\ref{tab:sch_app} called Schwarz's table.
As $\sigma=\tfrac{1}{2}$ only items $1$, $2$, $4$, $6$, $9$, or $14$
of the Table~\ref{tab:sch_app} are allowed. We analyse them case by case.
\paragraph{Case 1.}
\begin{itemize}
\item $\pm\rho=1/2+s$, for a certain $s\in \Z$, then $k=-2(mp+1)$ for
  a certain $p\in \Z$.  In this case $\tau$ is an arbitrary number, so
  $\lambda$ is arbitrary.
\item $\pm\tau=1/2+p$, for a certain $ p \in \Z$, then
  $\lambda\in\scI_2(k,m)$.  In this case $\rho$-arbitrary, and thus
  $k$ can be arbitrary.
\end{itemize}
\paragraph{Case 2.}
In this case $\pm\tau=1/3+p$, for a certain $ p\in \Z$, and
$\pm \rho=1/3+s$, for a certain $ s\in \Z$. The first condition
implies that $\lambda \in \scI_{3}(k,m)$. If the second condition is
fulfilled, then
\begin{equation}
  \label{eq:7}
  k = 2(m p-1) \pm \frac{1}{3}m. 
\end{equation} 
\paragraph{Case 4.}
We have two possibilities:
\begin{itemize}
\item if $\pm\rho=1/3+s$, and $\pm\tau=1/4+p$ for certain $s,p\in \Z$,
  then $k$ is given by~\eqref{eq:7}, and $\lambda\in\scI_4(k,m)$.
\item if $\pm\rho=1/4+s$, and $\pm\tau=1/3+p$, for certain
  $p, s\in \Z$, then
  \begin{equation}
    \label{eq:8}
    k = 2(m p-1) \pm \frac{1}{2}m ,
  \end{equation}
  and $\lambda\in\scI_{3}(k,m)$.
\end{itemize}
\paragraph{Case 6.}
\begin{itemize}
\item If $\pm\rho=1/3+s$ and $\pm\tau=1/5+p$, for some $s,p\in \Z$,
  then $k$ is given by ~\eqref{eq:7} and $\lambda\in\scI_5(k,m)$.
\item If $\pm\rho=1/5+s$ and $\pm\tau=1/3+p$, for some $s,p\in \Z$,
  then
  \begin{equation}
    \label{eq:9}
    k = 2(m p-1) \pm \frac{3}{5}m ,
  \end{equation}
  and $\lambda\in\scI_3(k,m)$.
\end{itemize}
\paragraph{Case 9.}
\begin{itemize}
\item If $\pm\rho=2/5+s$, and $ \pm\tau=1/5+p$, for some $s, p\in \Z$,
  then
  \begin{equation}
    \label{eq:10}
    k = 2(m p-1) \pm \frac{1}{5}m ,
  \end{equation}
  and $\lambda\in\scI_5(k,m)$.
   
\item If $\pm\rho=1/5+s$, and $\tau=2/5+p$, for some $s, p\in \Z$,
  then $k$ is given by~\eqref{eq:9} and $\lambda\in\scI_6(k,m)$.
\end{itemize}
\paragraph{Case 14.}
\begin{itemize}
\item If $\pm\rho=2/5+s$, and $\tau=1/3+p$, for some $ s, p\in \Z$,
  then $k$ is given by~\eqref{eq:10} and $\lambda\in\scI_3(k,m)$.
\item If $\pm\rho=1/3+s$, and $\pm\tau=2/5+p$, for some $s, p\in \Z$,
  then $k$ is given by~\eqref{eq:7} and $\lambda\in\scI_6(k,m)$.
\end{itemize}
Now, collecting all items with the same form of $k$ and taking into
account that it has to be an integer we obtain
Table~\ref{tab:integrability_table}.

\section{Application of Theorem~\ref{th:1}} 
Here we consider two examples of  Hamilton functions of the form~\eqref{eq:m_h}.
\paragraph{Example 1.}
First of all we check separable cases for Hamiltonian~\eqref{eq:m_h}.
Its Hamilton-Jacobi equation takes the form
\begin{equation}
  \label{eq:ham_jacobi}
  \frac{1}{2}r^{m-k}\left[\left(\pder{S}{r}\right)^2+r^{-2}\left(\pder{S}{\varphi}\right)^2\right]+r^mU(\varphi)=E,
\end{equation}
where $S=S(r,\varphi)$ is Hamilton's characteristic function. We look
for $S$ postulating its additive form
\[
S=S_r(r)+S_\varphi(\varphi).
\]
Then we can rewrite~\eqref{eq:ham_jacobi} in the following way
\begin{equation}
  \label{eq:ham_jacobi_2}
  r^{-k}\left(\oder{S_r}{r}\right)^2+r^{-(k+2)}\left(\oder{S_\varphi}{\varphi}\right)^2+2U(\varphi)=2r^{-m}E.
\end{equation}
  This equation separates when $k=-2$ and then we obtain
\begin{equation}
  r^2\left(\oder{S_r}{r}\right)^2-2r^{-m}E=\alpha,\qquad \left(\oder{S_\varphi}{\varphi}\right)^2+2U(\varphi)=-\alpha,
\end{equation}
where $\alpha$ is a separation constant. 
So, in this case the Hamiltonian of the  system  has the form 
\begin{equation}
  \label{eq:ham_ex_1}
  H=\frac{1}{2}r^{m+2}\left(p_r^2+\frac{p_\varphi^2}{r^2}\right)+r^mU(\varphi),
\end{equation}
and it  is integrable  with the following additional first integral
\[
  G=\frac{p_\varphi^2}{2}+U(\varphi).
\]
Let us note that the case   $k=-2$  is contained in the first item of  Table~\ref{tab:integrability_table}.
\paragraph{Example 2.}
Now we consider the Hamilton function
\begin{equation}
  \label{eq:m_hex}
  H=\frac{1}{2}r^{m-k}\left(p_r^2+\frac{p_\varphi^2}{r^2}\right)-r^m\cos\varphi.
\end{equation}
It has the form~\eqref{eq:m_h} with $U(\varphi)=-\cos\varphi$. As
$U'(\varphi)=\sin\varphi$, we take $\varphi_0=0$. Since $U''(0)/U(0)=-1$, we have  $\lambda=(k-1)/k$.
Comparing this value with  forms of $\lambda$ in sets $\scI_j(k,m)$ for
$j=0,\ldots,6$ we obtain the following conditions:
\begin{subequations}
\label{eq:condJ}
\begin{itemize}
\item if $\lambda\in\scI_0(k,m)$, then $ 2m^2p^2+(k+2)mp+1=0$, and this implies
  that 
\begin{equation}
\left[ 4mp +k +2\right]^2=k^2+4k-4,
\label{eq:cond0}
\end{equation} 
\item if 
$\lambda\in\scI_1(k,m)$, then $ m^2p^2-(k+2)mp+2=0$, and this implies that 
\begin{equation}
\left[ 2(mp-1) -k \right]^2=k^2+4k-4,
\label{eq:cond1}
\end{equation}
 \item if $\lambda\in\scI_2(k,m) $, then 
\begin{equation}
\left[m(2p+1)\right]^2=k^2+4k-4,\label{eq:cond2}
\end{equation}
\item if $\lambda\in\scI_3(k,m)$, then 
\begin{equation}
\left[2m(3p+1)\right]^2=9(k^2+4k-4),\label{eq:cond3}
\end{equation}
\item if $ \lambda\in\scI_4(k,m)$, then 
\begin{equation}
\left[m(4p+1)\right]^2=4(k^2+4k-4),\label{eq:cond4}
\end{equation}
\item if $\lambda\in\scI_5(k,m)$, then 
\begin{equation}
\left[2m(5p+1)\right]^2=25(k^2+4k-4),\label{eq:cond5}
\end{equation}
\item if $\lambda\in\scI_6(k,m)$, then  
\begin{equation}
\left[2m(5p+2)\right]^2=25(k^2+4k-4).\label{eq:cond6}
\end{equation}
\end{itemize}
\end{subequations}
It is easy to see that if  one of the above conditions is fulfilled, then 
\begin{equation}
k^2+4k-4=q^2,
\label{eq:condo}
\end{equation}
for a certain $q\in\Z$. Rewriting this equality  in the form 
\[
(k+2+q)(k+2-q)=8
\] 
and considering all decompositions of
$8=(\pm 1)\cdot (\pm 8)=(\pm 2)\cdot (\pm 4)=(\pm 4)\cdot (\pm 2)=(\pm 8)\cdot
(\pm 1)$,
we obtain that $k\in\{-5,1\}$. With these values of $k$ one can easily find that
$\lambda=(k-1)/k\in \scI_0(k,m)$ iff $m\in\{-1,1\}$. Hence, we have the
following four cases with the following $m$, $k$ and $l=m-k$:
\begin{equation}
\begin{split}
 &1.\quad\, m=1,\qquad k=-5, \quad\, l=6,\\
&2.\quad\,m=-1,\quad\, k=1, \qquad  l=-2,\\
&3.\quad\,m=1,\qquad k =1, \qquad  l=0,\\
&4.\quad\,m=-1,\quad\, k = -5, \quad\, l=4,
\end{split}
\label{eq:dup1}
\end{equation}
Similarly, if $\lambda\in (k-1)/k\in \scI_1$ with  $k\in\{-5,1\}$, then 
$m\in\{-2,-1,1, 2\}$. Thus, besides  cases~\eqref{eq:dup1} we have additionally
the following ones
\begin{equation}
\begin{split}
 &5.\quad\,m=2,\qquad k =1,\qquad l=1,\\
&6.\quad\,m=-2,\quad\, k= 1, \qquad  l=-3,\\
&7.\quad\,m=2,\qquad k=-5,\quad l=7,\\
&8.\quad\,m=-2,\quad\, k=-5, \quad\,  l=3.
\end{split}
\label{eq:dup2}
\end{equation}
Now, we  show that there are no other  cases when the necessary conditions for
the integrability given by Theorem~\ref{th:1}    in items 2-6 of Table~\ref{tab:integrability_table} are satisfied. In fact, for both values $k\in\{-5,1\}$
we have 
$k^2+4k-4=1$, thus the right-hand sides of conditions \eqref{eq:condJ} are given
explicitly. 
As the left- and the right-hand sides of equations \eqref{eq:cond3}, \eqref{eq:cond5}
and \eqref{eq:cond6} have different parities these equalities cannot hold.
 Equalities \eqref{eq:cond2} and \eqref{eq:cond4} are fulfilled   only for $m=\pm1$ or $m=\pm2$ and $p=0$, respectively, but these
 values are already  given in the above listed  cases.

Surprisingly all  cases \eqref{eq:dup1} are integrable and in fact  superintegrable.
\paragraph{Case 1.}
In this case we have the Hamiltonian of the following form
\begin{equation}
\label{eq:ham_1}
H=\frac{1}{2}r^6\left(p_r^2+\frac{p_\varphi^2}{r^2}\right)-r \cos\varphi.
\end{equation}
This system has two additional, functionally independent first integrals of the second order  in momenta
\begin{equation}
\begin{split}
F_1&:=r^2p_\varphi^2\cos(2\varphi) -r^3p_rp_\varphi\sin(2\varphi)+r^{-1}\sin\varphi\sin(2\varphi),\\
F_2&:=r^2p_\varphi^2\sin(2\varphi)+r^3p_rp_\varphi\cos(2\varphi)-r^{-1}\sin\varphi\cos(2\varphi).
\end{split}
\end{equation}
\paragraph{Case 2.}
We have the following Hamiltonian
\begin{equation}
\label{eq:ham_2}
H=\frac{1}{2}r^{-2}\left(p_r^2+\frac{p_\varphi^2}{r^2}\right)-r^{-1} \cos\varphi,
\end{equation}
and  two additional functionally independent first integrals take the form
\begin{equation}
\begin{split}
F_1&:=r^{-2}p_\varphi^2\cos(2\varphi)+r^{-1}p_rp_\varphi\sin(2\varphi)+r\sin\varphi\sin(2\varphi), \\ 
F_2&:=-r^{-2}p_\varphi^2\sin(2\varphi)+r^{-1}p_rp_\varphi\cos(2\varphi)+r\sin\varphi\cos(2\varphi).
\end{split}
\end{equation}
\paragraph{Case 3.}
Hamilton function and additional first integrals are the following
\begin{equation}
\label{eq:ham_3}
\begin{split}
H&=\frac{1}{2}\left(p_r^2+\frac{p_\varphi^2}{r^2}\right)-r\cos\varphi,\\
F_1&:=r^{-1}p_\varphi^2\cos\varphi+p_rp_\varphi\sin\varphi+\frac{1}{2}r^2\sin^2\varphi,\\
F_2&:=\left(p_r^2-r^{-2}p_\varphi^2\right)\cos\varphi\sin\varphi+r^{-1}p_rp_\varphi\cos(2\varphi)-r\sin\varphi.
\end{split}
\end{equation}
\paragraph{Case 4.}
In this case we have respectively:
\begin{equation}
\label{eq:ham_4}
\begin{split}
H&=\frac{1}{2}r^4\left(p_r^2+\frac{p_\varphi^2}{r^2}\right)-r^{-1}\cos\varphi,\\
F_1&:=rp_\varphi^2\cos\varphi-r^2p_rp_\varphi\sin\varphi+\frac{1}{2}r^{-2}\sin^2\varphi, \\
F_2&:=r^4\left(p_r^2-r^{-2}p_\varphi^2\right)\cos\varphi\sin\varphi-r^3p_rp_\varphi\cos(2\varphi)-r^{-1}\sin\varphi.
\end{split}
\end{equation}
\begin{figure}[h!]
  \centering \subfigure[section plane $r=1$  with coordinates $(\varphi,p_{\varphi})$]{
    \includegraphics[width=0.46\textwidth]{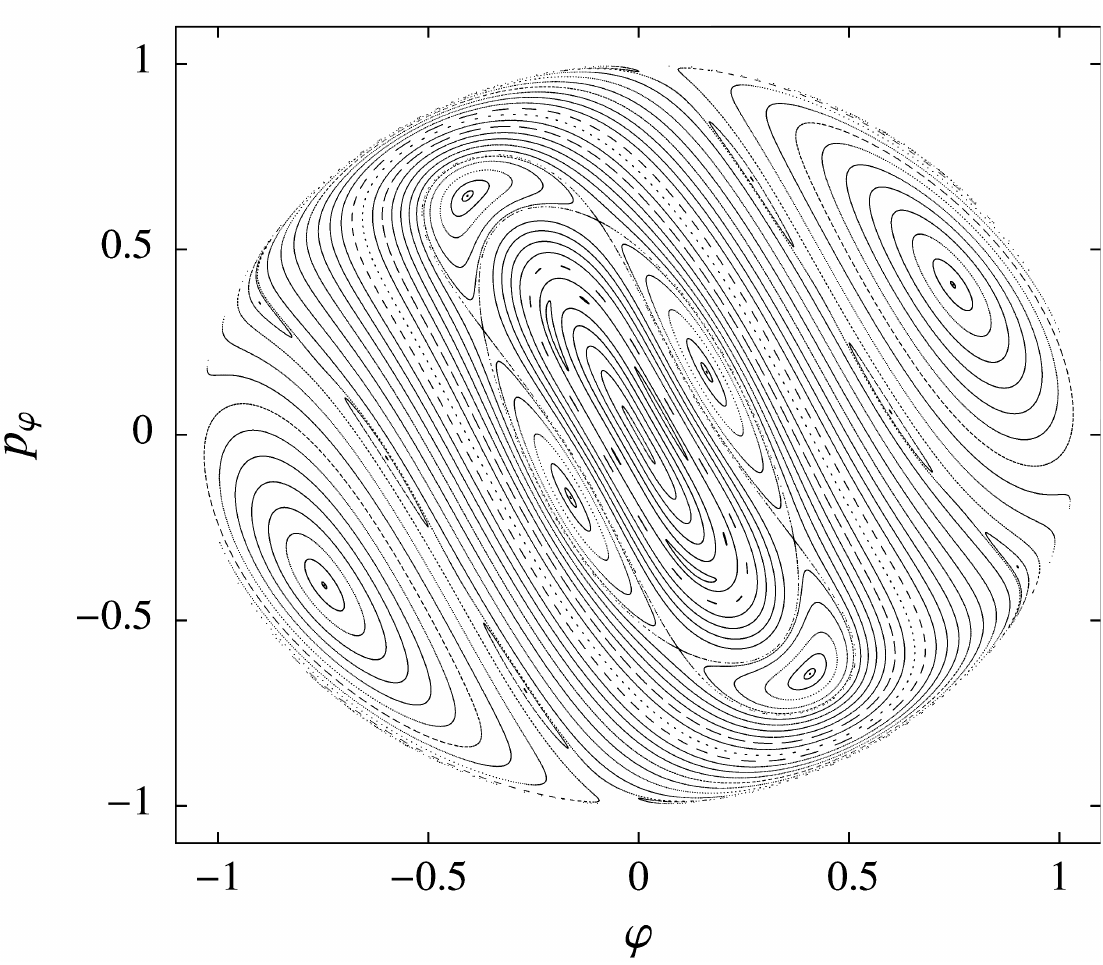}
  }\quad\subfigure[magnification of region around unstable periodic solution  \label{fig:mag1}]{
    \includegraphics[width=0.46\textwidth]{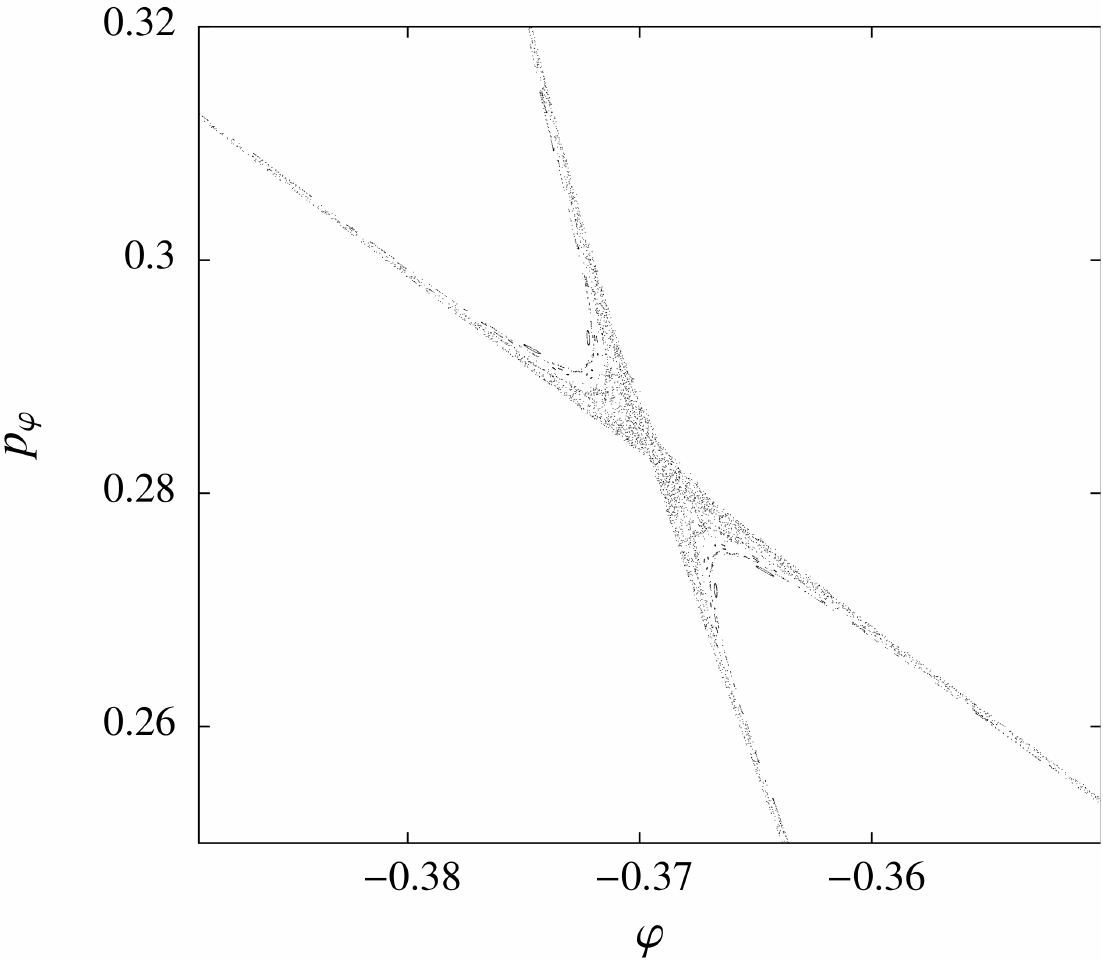}
  }
  \caption{\small Poincar\'e cross sections on energy level $E=-0.5$ for Hamiltonian system given by~\eqref{eq:m_hex} with  $ m=-2, k=1$ corresponding to Case 6 \label{case6}}
  \end{figure}   
  
 \begin{figure}[h!] 
    \centering \subfigure[section plane $r=1$  with coordinates $(\varphi,p_{\varphi})$]{
    \includegraphics[width=0.46\textwidth]{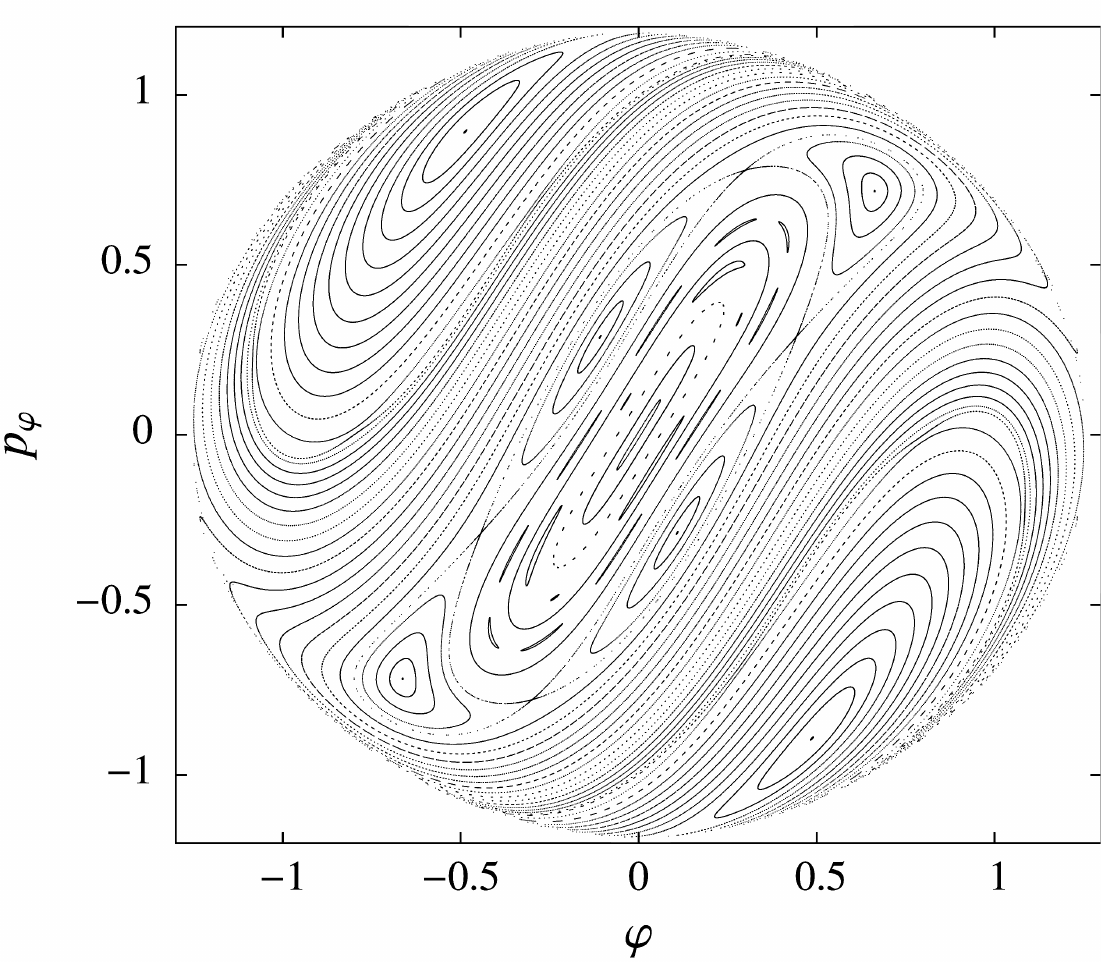}
  } \quad\subfigure[magnification of region around unstable periodic solution\label{fig:mag2}]{
    \includegraphics[width=0.46\textwidth]{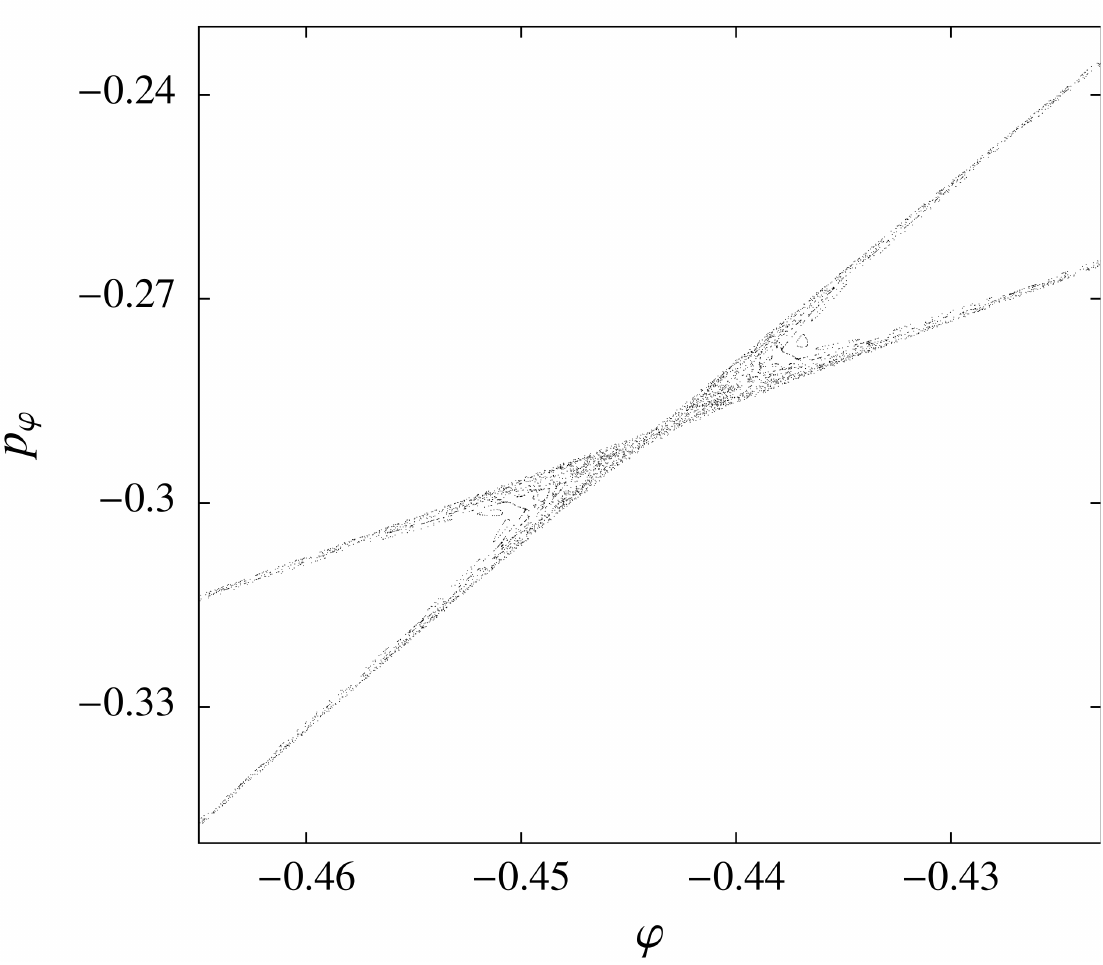}
  }
  \caption{\small Poincar\'e cross sections on energy level $E=-0.3$ for Hamiltonian system given by~\eqref{eq:m_hex} with  $ m=2,k=-5$  corresponding to Case 7 \label{case7}}
\end{figure}

\begin{figure}[h!]
  \centering \subfigure[section plane $r=1$  with coordinates $(\varphi,p_{\varphi})$]{
    \includegraphics[width=0.455\textwidth]{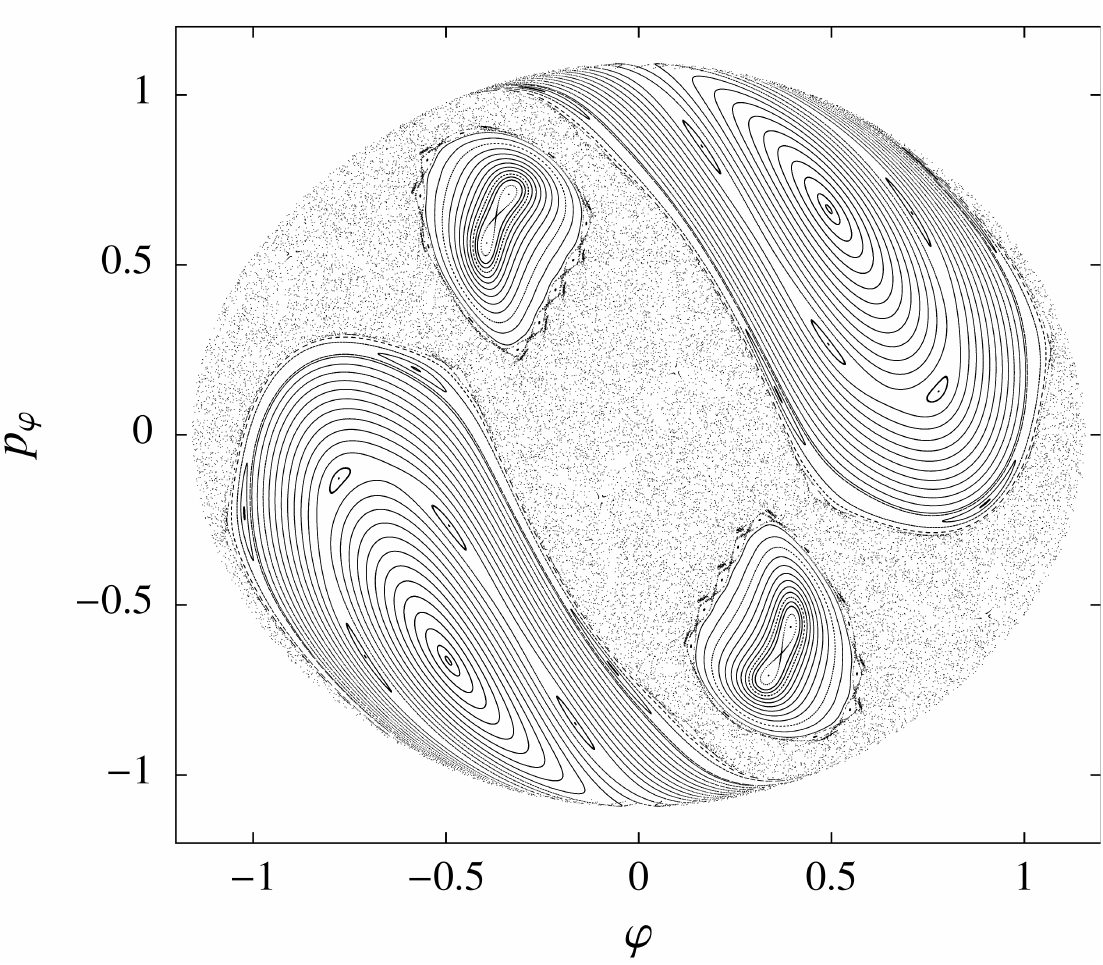}
  } \subfigure[section plane $\varphi=0$ with coordinates $(r,p_{r})$]{
    \includegraphics[width=0.455\textwidth]{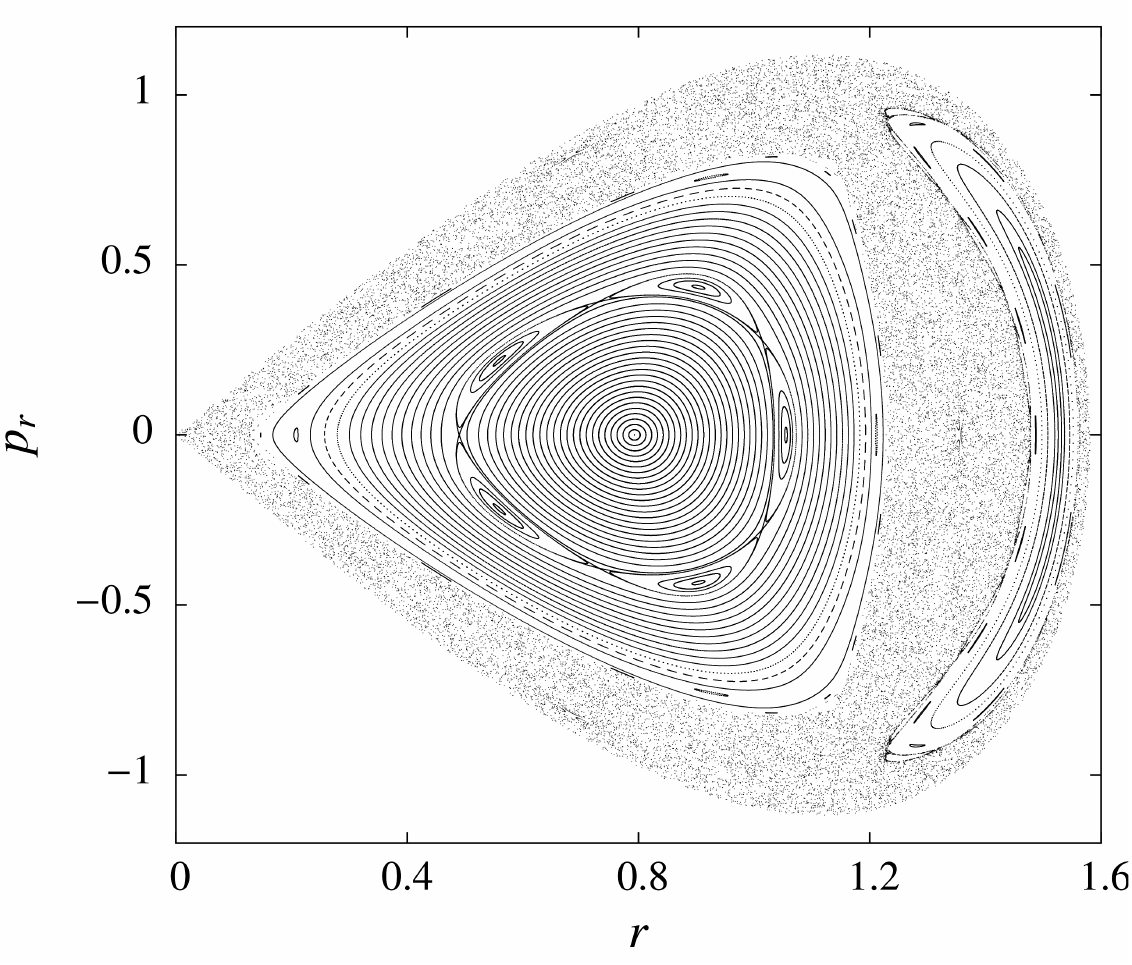}
  }
  \caption{\small Poincar\'e cross sections on energy level $E=-0.5$ for Hamiltonian system given by~\eqref{eq:m_hex} with $ m=-2, k=2$ \label{poincrossmn1}}
   \end{figure}   
  
 \begin{figure}[h!]  
  \centering \subfigure[section plane $r=1$  with coordinates $(\varphi,p_{\varphi})$]{
    \includegraphics[width=0.455\textwidth]{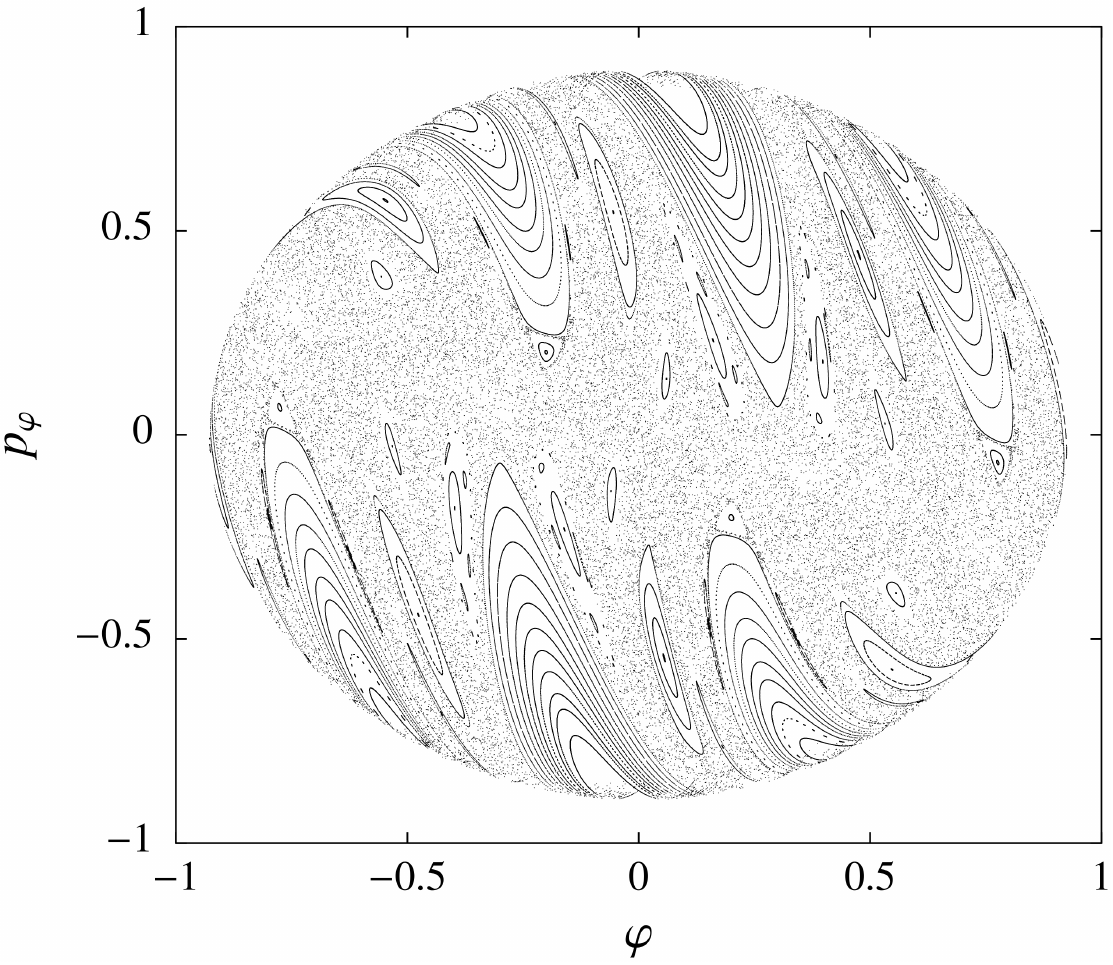}
  } \subfigure[section plane $\varphi=0$ with coordinates $(r,p_{r})$ ]{
    \includegraphics[width=0.46\textwidth]{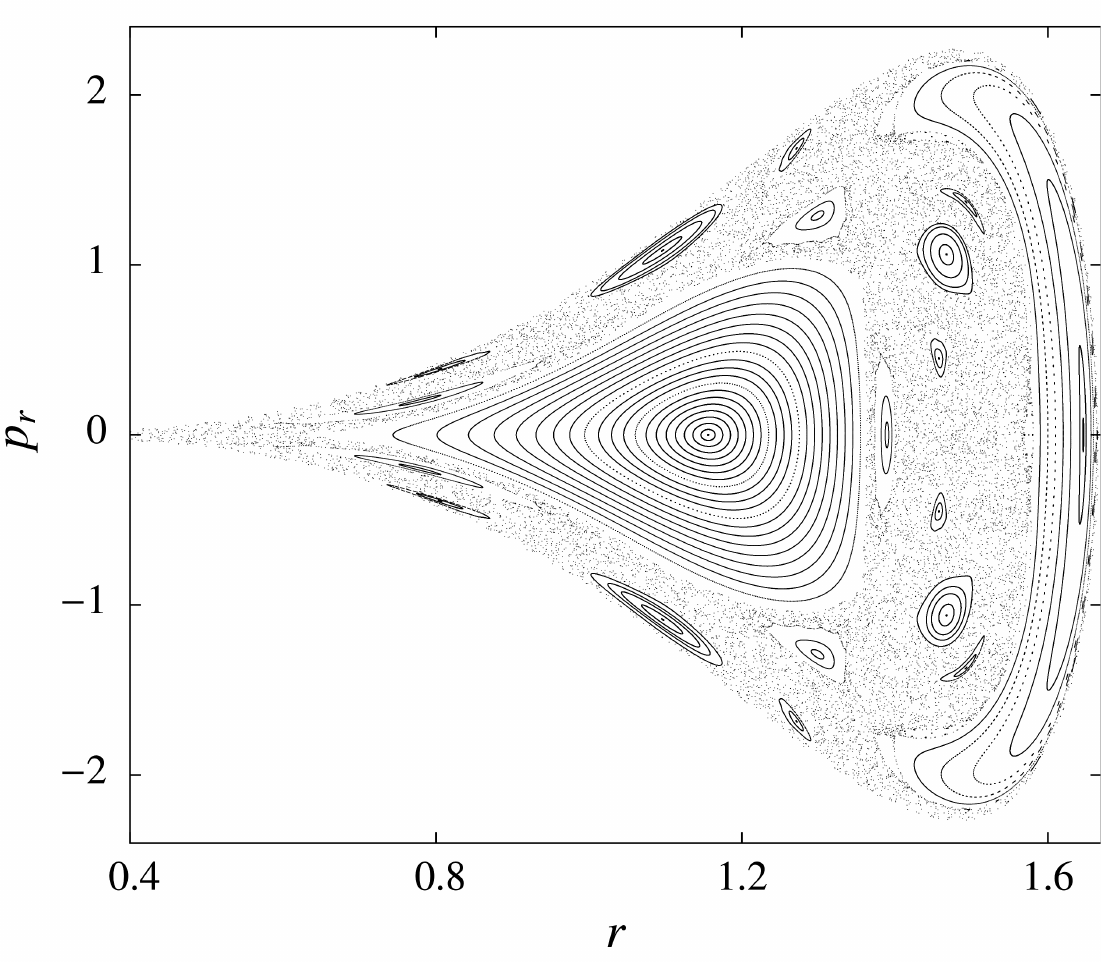}
  }
  \caption{\small Poincar\'e cross sections on energy level $E=-0.6$ for   Hamiltonian system given by~\eqref{eq:m_hex} with $ m=-1,k=8$ \label{poincrossmn2}}
  \end{figure}
  \begin{figure}[h!]
  \centering \subfigure[section plane $r=1$  with coordinates $(\varphi,p_{\varphi})$]{
    \includegraphics[width=0.46\textwidth]{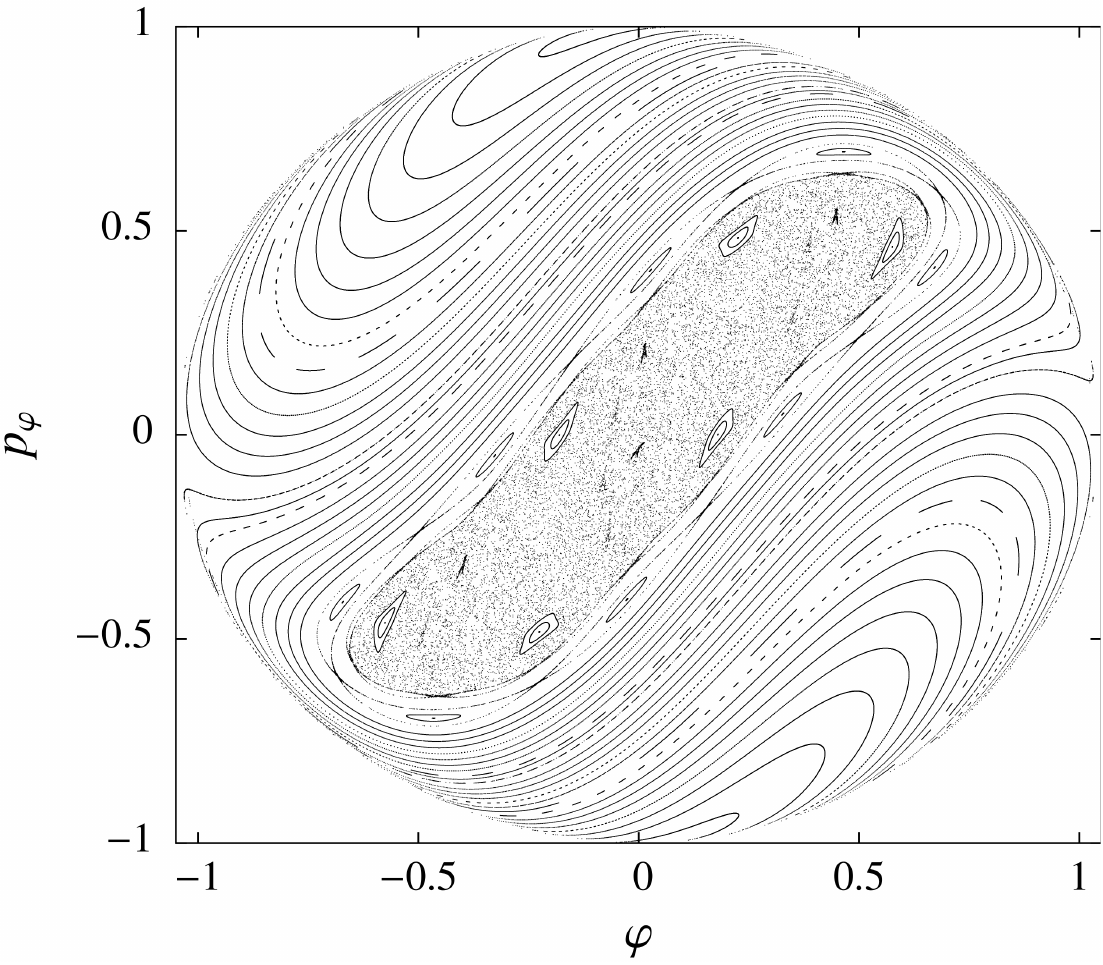}
  } \subfigure[section plane $\varphi=0$ with coordinates $(r,p_{r})$ ]{
    \includegraphics[width=0.46\textwidth]{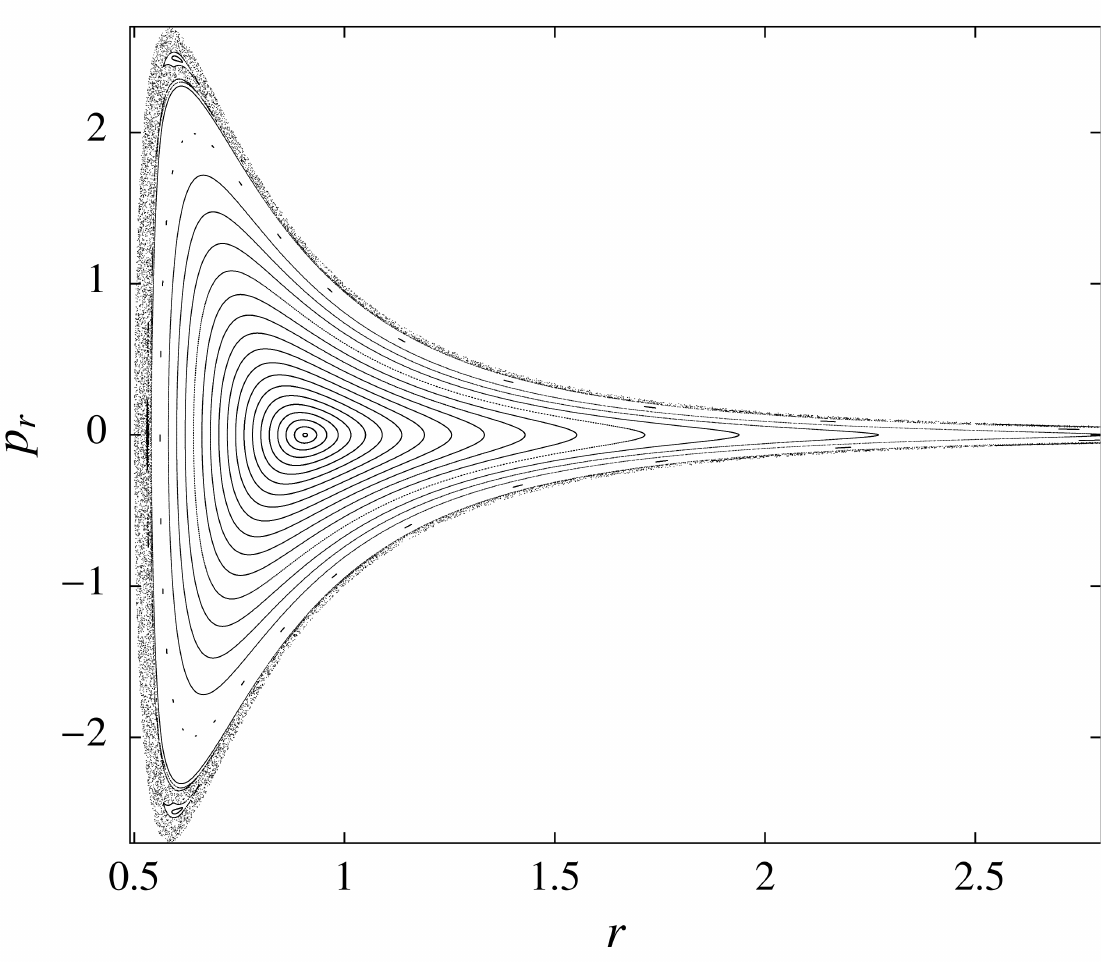}
  }
  \caption{\small Poincar\'e cross sections on energy level $E=-0.5$ for  Hamiltonian system given by~\eqref{eq:m_hex} with $ m=1, k=-6$ \label{poincrossmn3}}
\end{figure}

In cases with parameters given in \eqref{eq:dup2} we have integrable as well as non-integrable systems. Namely 
 cases $5$ and $8$ are integrable.
\paragraph{Case 5.}
\begin{equation}
\begin{split}
H&=\frac{1}{2}r\left(p_r^2+\frac{p_\varphi^2}{r^2}\right)-r^{2}\cos\varphi,\\
F&:=r^{-1}(p_\varphi^2-r^2p_r^2)\cos\varphi+r^2(1+\cos^2\varphi)+2p_r p_{\varphi}\sin\varphi.
\end{split}
\end{equation}
\paragraph{Case 8.}
\begin{equation}
\begin{split}
H&=\frac{1}{2}r^{3}\left(p_r^2+\frac{p_\varphi^2}{r^2}\right)-r^{-2}\cos\varphi,\\
F&:=r(p_\varphi^2-r^2p_r^2)\cos\varphi+r^{-2}(1+\cos^2\varphi)-2r^2p_rp_\varphi\sin\varphi.
\end{split}
\end{equation}

However the Poincar\'e sections for Hamiltonian system~\eqref{eq:m_hex}  with parameters given in Cases $6$ and $7$ in \eqref{eq:dup2} show chaotic area, see Figures~\ref{case6}-\ref{case7}.  Chaos appears just in  vicinities of  unstable periodic solutions visible on magnifications~\ref{fig:mag1} and~\ref{fig:mag2}. 

Cases from the first  item of Table~\ref{tab:integrability_table} are generically non-integrable, see Poincar\'e sections on Figures~\ref{poincrossmn1}-\ref{poincrossmn3} showing large chaotic regions. However in order to prove the non-integrability of  Hamiltonians with $m$ and $k$ given by this item higher order variational equations must be used. 

\section*{Acknowledgement}
The work  has been supported by grant No. DEC-2013/09/B/ST1/04130 of National Science Centre of Poland. The authors are grateful to the anonymous referee for providing helpful corrections and suggestions improving the text.
\appendix
\section{Gauss hypergeometric equation} 
\label{sec:hyper}
The  Riemann $P$ equation  is the most general
second order differential equation with three regular singularities  \cite{Whittaker:35::,Kristensson:12::}.
If we place using homography these singularities at $z\in\{0,1,\infty\}$,
then it has the form
\begin{equation}
\label{eq:riemann}
\begin{split} 
\dfrac{\mathrm{d}^2\eta}{\mathrm{d}z^2}&+\left(\dfrac{1-\alpha-\alpha'}{z}+ 
\dfrac{1-\gamma-\gamma'}{z-1}\right)\dfrac{\mathrm{d}\eta}{\mathrm{d}z}+ 
\left(\dfrac{\alpha\alpha'}{z^2}+\dfrac{\gamma\gamma'}{(z-1)^2}+ 
\dfrac{\beta\beta'-\alpha\alpha'-\gamma\gamma'}{z(z-1)}\right)\eta=0, 
\end{split} 
\end{equation}
where $(\alpha,\alpha')$, $(\gamma,\gamma')$ and $(\beta,\beta')$ are the
exponents at the respective singular points. These exponents satisfy
the Fuchs relation
\[
\alpha+\alpha'+\gamma+\gamma'+\beta+\beta'=1.
\]
We denote the differences of exponents by
\[
\rho=\alpha-\alpha',\qquad\sigma=\beta-\beta',\qquad \tau=\gamma-\gamma'.
\]
In particular the Gauss hypergeometric equation
\begin{equation}\label{eq:gausss}
\dfrac{\mathrm{d}^2\eta}{\mathrm{d}z^2}+\left(\frac{(\alpha+\beta+1)z-\gamma}{z(z-1)}\right)\dfrac{\mathrm{d}\eta}{\mathrm{d}z}+\frac{\alpha\beta}{z(z-1)}\eta=0,
\end{equation}
where $\alpha,\beta,\gamma$ are parameters with the exponent differences given by
\[\rho=1-\gamma,\qquad \sigma=\gamma-\alpha-\beta,\qquad \tau=\beta-\alpha.\]
is a special form of Riemann $P$ equation.
Necessary and sufficient conditions for solvability of the identity
component of the differential Galois group of \eqref{eq:riemann} and \eqref{eq:gausss} are
given by the following theorem due to Kimura  \cite{Kimura:69::}.
\begin{theorem} 
\label{th:Kimura}
  The identity component of the differential Galois group of the Riemann $P$ equation \eqref{eq:riemann} is solvable iff
\begin{description} 
\item[A.] at least one of the four numbers $\rho+\sigma+\tau$,
  $-\rho+\sigma+\tau$, $\rho+\sigma-\tau$, $\rho-\sigma+\tau$ is an odd
  integer, or
\item[B.] the numbers $\rho$ or $-\rho$  and
  $\sigma$ or $-\sigma$ and $\tau$ or $-\tau$ belong (in an arbitrary order) to some of
  appropriate fifteen families forming the so-called Schwarz's Table \ref{tab:sch_app}.
\begin{table}[h]
 \begin{tabular}{lllll}
    \toprule
        \text{1}&$1/2+r$&$1/2+s$&arbitrary complex number&\\
        \text{2}&$1/2+r$&$1/3+s$&$1/3+p$&\\
        \text{3}&$2/3+r$&$1/3+s$&$1/3+p$&$r+s+p$ even\\
        \text{4}&$1/2+r$&$1/3+s$&$1/4+p$&\\
        \text{5}&$2/3+r$&$1/4+s$&$1/4+p$&$r+s+p$ even\\
        \text{6}&$1/2+r$&$1/3+s$&$1/5+p$&\\
        \text{7}&$2/5+r$&$1/3+s$&$1/3+p$&$r+s+p$ even\\
        \text{8}&$2/3+r$&$1/5+s$&$1/5+p$&$r+s+p$ even\\
        \text{9}&$1/2+r$&$2/5+s$&$1/5+p$&\\
        \text{10}&$3/5+r$&$1/3+s$&$1/5+p$&$r+s+p$ even\\
        \text{11}&$2/5+r$&$2/5+s$&$2/5+p$&$r+s+p$ even\\
        \text{12}&$2/3+r$&$1/3+s$&$1/5+p$&$r+s+p$ even\\
        \text{13}&$4/5+r$&$1/5+s$&$1/5+q$&$r+s+p$ even\\
        \text{14}&$1/2+r$&$2/5+s$&$1/3+p$& \\
        \text{15}&$3/5+r$&$2/5+s$&$1/3+p$&$r+s+p$ even\\
  \bottomrule
  \end{tabular}
   \caption{The Schwarz's table. Here $r,s,p\in\Z$ \label{tab:sch_app}}
\end{table}
\end{description}
\end{theorem}

\end{document}